\documentclass[12pt]{iopart}
\usepackage{graphicx}% Include figure files
\usepackage{dcolumn}% Align table columns on decimal point
\usepackage{bm}% bold math
\usepackage[UKenglish]{babel}
\usepackage{amsmath}
\usepackage{url}

\begin{document}

\title[]{Universality of weakly bound dimers and Efimov trimers close to Li-Cs Feshbach resonances}

\author{J. Ulmanis, S. H\"afner, R. Pires, E. D. Kuhnle, M. Weidem\"uller\footnote{Also at: Hefei National Laboratory for Physical Sciences at the Microscale, University of Science and Technology of China, Hefei, Anhui 230026, People's Republic of China}}
\address{Physikalisches Institut, Universit\"at Heidelberg, Im Neuenheimer Feld 226, 69120 Heidelberg, Germany} 
\ead{weidemueller@uni-heidelberg.de}

\author{E. Tiemann}
\address{Institut f\"{u}r Quantenoptik, Leibniz Universit\"{a}t Hannover, Welfengarten 1, 30167 Hannover, Germany}


\begin{abstract}
	We study the interspecies scattering properties of ultracold Li-Cs mixtures in their two energetically lowest spin channels in the magnetic field range between 800 G and 1000 G. Close to two broad Feshbach resonances we create weakly bound LiCs dimers by radio-frequency association and measure the dependence of the binding energy on the external magnetic field strength. Based on the binding energies and complementary atom loss spectroscopy of three other Li-Cs s-wave Feshbach resonances we construct precise molecular singlet and triplet electronic ground state potentials using a coupled-channels calculation. We extract the Li-Cs interspecies scattering length as a function of the external field and obtain almost a ten-fold improvement in the precision of the values for the pole positions and widths of the s-wave Li-Cs Feshbach resonances as compared to our previous work [Pires \textit{et al.}, Phys. Rev. Lett. \textbf{112}, 250404 (2014)]. We discuss implications on the Efimov scenario and the universal geometric scaling for LiCsCs trimers.

\end{abstract}


\maketitle

%\tableofcontents

\section{Introduction}

Universality in few-body systems has been one of the major topics in the ultracold quantum gases for the last decade \cite{Braaten2006,Ferlaino2011,Wang2013}. Its success can be ascribed to the existence of magnetic interparticle scattering resonances, called Feshbach resonances (FR), at which a two-body bound state crosses the scattering threshold. These resonances are routinely employed to tune the interaction strength between the colliding particles \cite{chin2010} and to produce weakly bound dimers by ramping up or down the external magnetic field \cite{Koehler2006}. They can be used to explore intriguing topics in few-body physics, for example, the realization of Efimov's scenario \cite{Braaten2006,Wang2013}. Its hallmark is the existence of a geometrical series of weakly bound three-body states that exhibits the universal scaling law. The energy of the next bound trimer can be found by multiplying the binding energy of the previous one with a constant factor. However, the ability to produce and study these trimers as well as underlying universal principles, on which the behavior of such exotic systems is based, relies on precise knowledge of the properties of the particular FR. 

The central quantity that governs an ultracold collision process, and therefore most of the physics at such temperatures, is the two-body s-wave scattering length $ a $. The inelastic three-body scattering rate near a FR scales as $ a^4 $, resulting in magnetic field dependent atom losses that can be used to map out how $ a $ depends on the external field. During the last decade, atom loss spectroscopy in combination with theoretical models has become a standard tool in the field of ultracold gases \cite{chin2010}. These methods can give an excellent representation of the FR spectrum, however typically not all of the observed losses can be unambiguously attributed to an increasing two-body scattering length. Especially when $ a $ becomes large, not only immediate loss of three atoms from the trap, but also other processes, for example, weakly-bound dimer formation and subsequent secondary losses may occur. This may lead to shifts and asymmetric broadening of the loss signals \cite{Weber2008,Machtey2012,Zhang2011c, Khramov2012} and thus weakening the relation to the functional dependence of the scattering length alone. 

More accurate mapping can be obtained by going further than a simple atom loss spectrum. The most precise scattering length measurements up to date can be obtained by direct radio-frequency (rf) \cite{Klempt2008,Regal2003,Bartenstein2005,Ospelkaus2006,Wu2012,Zuern2013} and magnetic field modulation \cite{Weber2008,Thompson2005,Papp2006,Lange2009,Gross2010,Berninger2013,Dyke2013} spectroscopy of the least-bound molecular states. Since a FR intrinsically originates from the coupling of the scattering channel with such a molecular state, its energy $ E $ in the vicinity of the FR can be connected to the scattering length through the relation $ E\sim a^{-2} $ \cite{chin2010,Gribakin1993}. By mapping the magnetic field dependence of the binding energy of this state, it is possible to study exclusively the two-body problem, and the extraction of $ a $ is less prone to systematic effects.

Here we explore the universal behaviour of weakly bound LiCs dimers and Efimov trimers close to Li-Cs FRs. We start by investigating the interspecies scattering properties of ultracold Li-Cs mixtures of the two energetically lowest LiCs spin channels in the magnetic field range between 800 G and 1000 G. We employ rf association and atom loss spectroscopy to precisely measure the positions of the LiCs s-wave FRs in this magnetic field range. Depending on the width of the resonance we separate them into two groups and use complementary approaches to determine their properties. For the broad resonances close to 843 G and 889 G we measure magnetic field dependent binding energies of weakly bound dimers through rf association. For the narrow resonances we employ atom loss spectroscopy. Due to their small width the resonance position can still be detected with high accuracy.
We use these measurements as an input for a coupled-channels (cc) calculation that allows us to construct accurate Li-Cs molecular potentials, from which scattering lengths, resonance positions and widths are determined. The obtained parameters agree well with the previous observations \cite{Repp2013,Tung2013} and recent extensive theoretical studies \cite{Pires2014a}, however they represent almost an order of magnitude improvement in precision and accuracy. Finally, we discuss the implications of these results to the recent observation of LiCsCs Efimov resonances \cite{Pires2014,Tung2014}. With the help of the new mapping of the scattering length in the vicinity of the 843 G LiCs FR we obtain refined Efimov scaling factors of 5.5(2) and 5.0(1.5) for the first and second Efimov period, respectively, where the first period slightly deviates from the universal Efimov scenario of 4.9, as predicted for the Li-Cs system with mass ratio of 22 \cite{DIncao2006a}.

\section{Radio-frequency association of LiCs Feshbach dimers}
The sample preparation scheme for the rf association measurements is similar to the one presented previously in Refs. \cite{Repp2013,Pires2014}. In brief, we prepare the Li-Cs mixture in a crossed optical dipole trap with standard laser-cooling techniques. Using degenerate Raman sideband cooling \cite{Kerman2000}  most of the Cs atoms are optically pumped and spin-polarized in the energetically lowest spin state $ \left|F=3,m_F=3\right\rangle $. After the initial cooling steps the Li atoms populate both energetically lowest spin states, namely $ \left|F=1/2,m_F=1/2\right\rangle $ and $ \left|F=1/2,m_F=-1/2\right\rangle $. The last forced evaporation ramp is performed at 920 G. At the end of the ramp one of the two Li spin components is selected by shining in a short, resonant light pulse that expels the other one from the trap. Finally, about  $ 3\cdot10^4 $ ($ 4\cdot10^4 $) atoms remain in the respective Cs (Li) spin channels with a temperature around 400 nK for each species. We measure the trapping frequencies $f_x, f_y, f_z$ of 11~Hz, 114~Hz,  123~Hz  (33~Hz, 275~Hz,  308~Hz) for Cs (Li) atoms, where the external magnetic field is parallel to the z axis. The total uncertainty of the applied magnetic field amounts to 16 mG (one standard deviation), resulting from long-term magnetic field drifts, residual field curvature along the long axis of the cigar-shaped trap and calibration uncertainties.

 \begin{figure}
 	\centering
 	\includegraphics[width=0.5\linewidth]{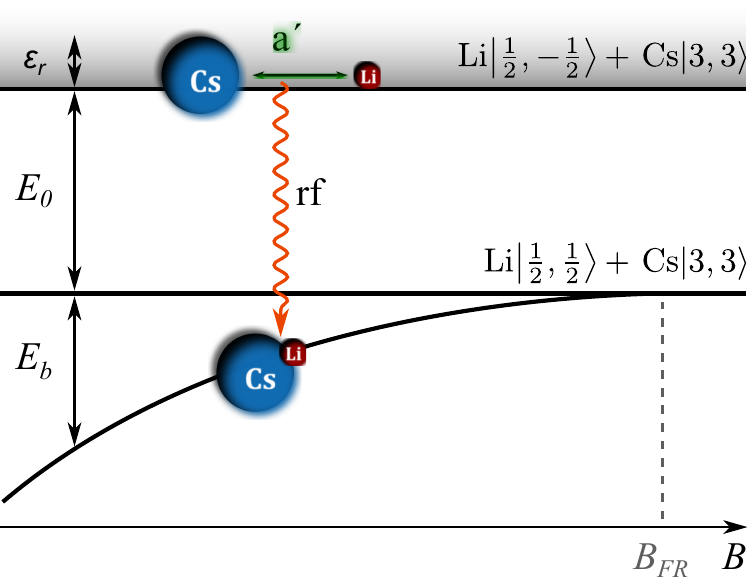}
 	\caption[RF association experiment]{Radio-frequency association of Li-Cs molecules. The mixture is initially prepared in the non-resonant scattering channel, here $ \mathrm{Li}\left|1/2,-1/2\right\rangle \oplus \mathrm{Cs}\left|3,3\right\rangle $, at a magnetic field close to the broad 843 G s-wave Feshbach resonance in the resonant scattering channel $ \mathrm{Li}\left|1/2,1/2\right\rangle \oplus \mathrm{Cs}\left|3,3\right\rangle $, which couples to the weakly bound molecular state under study.  Depending on the frequency of the rf driving field either free-free (with the energy $ E_0 $) or free-bound (with the energy $ E_0+E_b $) transition can be studied. An analogous scenario is implemented close to the second broad Li-Cs Feshbach resonance in the  $ \mathrm{Li}\left|1/2,-1/2\right\rangle \oplus \mathrm{Cs}\left|3,3\right\rangle $ scattering channel close to 889 G. In this case the mixture is initially prepared in the $ \mathrm{Li}\left|1/2,1/2\right\rangle \oplus \mathrm{Cs}\left|3,3\right\rangle $ channel.}
 	\label{fig:RF_experiment}
 \end{figure}

To associate the molecules we start with a mixture prepared in the non-resonant scattering channel at a variable magnetic field close to the FR in the resonant state (see Fig. \ref{fig:RF_experiment}). We drive the system with a rectangular rf pulse with the frequency $ E_{rf}/h $ that is close to the resonance frequency $ E_0/h $ between the two energetically lowest Li spin states. In order to determine the molecular binding energy $ E_b $, we scan the frequency of the applied rf field and observe the number of Li atoms that are left in the non-resonant state after the rf pulse.  

A typical loss spectrum is depicted in Fig. \ref{fig:843G_LossSpectrum}. Detuned from the free-free transition, which corresponds to a flip of Li nuclear spin, we observe an additional loss feature that originates from the association of LiCs Feshbach dimers (free-bound transition).  We also identify a similar loss signal at comparable values of detuning and amplitude in the remaining number of Cs atoms. To limit saturation effects, we experimentally optimize the power and length of the association pulse such that at most 30\% of atoms are lost at the end of the rf pulse. The optimized pulse length ranges from 0.5 s close to the FR and 7 s away from it.
 
\section{Theoretical analysis}

\subsection{LiCs dimer association spectrum}

\begin{figure}
\centering
\includegraphics[width=0.6\linewidth]{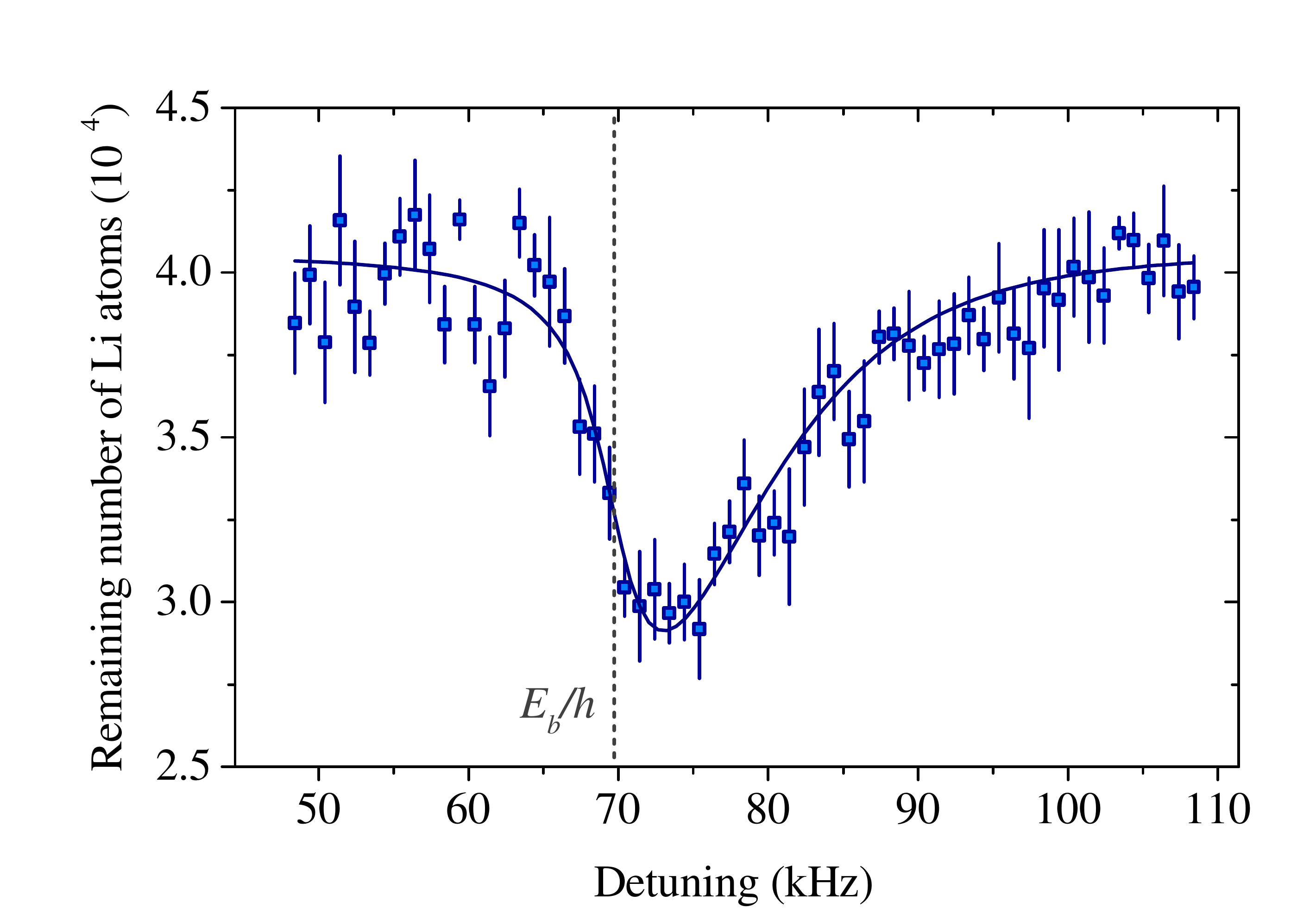}
\caption[Typical loss spectrum at 842.04G]{Remaining number of Li atoms after an rf association (free-bound) pulse of LiCs molecules at a magnetic field of 842.04 G and relative Li-Cs temperature of 400 nK. Each data point is an average of three measurements and the error bars represent the standard error. The binding energy is determined from the fit of Eq.~\ref{eq:rate_eq} to the data (solid line) and yields $ E_b/h=69.7(1.6) \mathrm{\ kHz}$ and $ \gamma=5(1) \mathrm{\ kHz}$ for a pulse length $ t=3 \mathrm{\ s} $. The vertical dashed line corresponds to the fitted binding energy.}
\label{fig:843G_LossSpectrum}
\end{figure}

We model the observed loss spectrum with the help of rate equations and the dimer binding energy dependent two-body association rate $ K_2^M $ \cite{Chin2005a,Klempt2008}. The long association pulse lengths and low molecule yield, which is below our detection limit, indicates that the dimer association rate is much smaller than their loss rate. Assuming a quasi-stationary state, in which each produced molecule immediately gets lost through atom-dimer collisions, the time dependent Li atom loss at a given magnetic field is governed by $ K_2^M $ and thus can be described through the following equation
\begin{equation}
N_{Li}=N_{Li}^{0}e^{-n_{Cs} K_2^M t},
\label{eq:rate_eq}
\end{equation} 
 where $ N_{Li}^{0} $ is the initial number of Li atoms in the non-resonant state, $ n_{Cs} $ denotes the density of the Cs gas cloud, $ t $ is the length of the applied rf pulse and $ K_2^M $ contains the functional form (see below) of the molecular association rate.
 Here we neglect single-body losses and assume constant $ n_{Cs} $, which is justified by the fact that we are working in a low saturation limit of the rf transition. By solving the full system of rate equations we estimate that this approximation, using the simple Eq.~\ref{eq:rate_eq},  may introduce a minor error on the fitted atom loss amplitude, which does not exceed 10 \%. Since the number of produced LiCs molecules at any given point through the experimental cycle is insignificant, we do not include the loss terms associated with the molecule-molecule recombination. 
 
 The two-body association rate $ K_2^M $ is determined by the energy-dependent wave-function overlap of the scattering atom pair with the final molecular state\cite{Chin2005a}. For a thermal ensemble it can be expressed as 
\begin{equation}
K_2^M(E_{rf})=C \int\limits_{0}^{\infty} h(\varepsilon_r)  F(\varepsilon_r,E_b)  L_{\gamma}(E_{rf},E_0+E_b+\varepsilon_r) d\varepsilon_r,
\label{eq:K2_Rate}
\end{equation}
 where $ h(\varepsilon_r)\propto e^{-\varepsilon_r/k_b T} $ is the number density of colliding atom pairs with relative energy $ \varepsilon_r $ and temperature $ T $, and
 \begin{equation}
 	F(\varepsilon_r,E_b) \propto \left(1-\sqrt{\frac{E_b}{E^\prime_b}}\right)^2 \frac{\sqrt{\varepsilon_r E_b} E^{\prime}_b}{(\varepsilon_r+E_b)^2(\varepsilon_r+E^\prime_b)}
 	\label{eq:FC_factor}
 \end{equation}
 is the energy normalized Franck-Condon density between the scattering wave function of a free Li-Cs atom pair and a bound Feshbach dimer with binding energy $ E_b $ \cite{Chin2005a,Klempt2008}. $ E^\prime_b $  is defined through the Li-Cs reduced mass $ \mu $ and the non-resonant channel scattering length $ a^\prime $ as $ E^\prime_b=\hbar^2/(2\mu a^{\prime 2}) $. The convolution of the spectroscopic line shape with the Lorentzian profile $ L_{\gamma}(E_{rf},E_0+E_b+\varepsilon_r) $ of width $ \gamma $ accounts for the strong collisional broadening, yielding an estimated lifetime of LiCs molecules in the mixture around $ 30 \ \mu\mathrm{s}$. The prefactor $ C $ contains all the numerical factors resulting from the integration of rate equations, and experimental parameters that affect the molecule production rate, but which are approximately constant for a given magnetic field, as well as species-dependent  atom-dimer inelastic collision rates\footnote{The prefactor $ C $ depends on $ n_{Cs} $ and $ n_{Li} $. For atom losses, which do not exceed $ \approx$ 30\% of the initial number of atoms, it does not change by more than 5 \%.}. It also accounts for uncertainties in the determination of the absolute gas densities, which, under realistic experimental conditions, can vary up to a factor of two due to systematic errors in measurements of the trap frequencies, temperature and the exact number of atoms.
 
\begin{figure}
 	\centering
 	\includegraphics[width=1\linewidth]{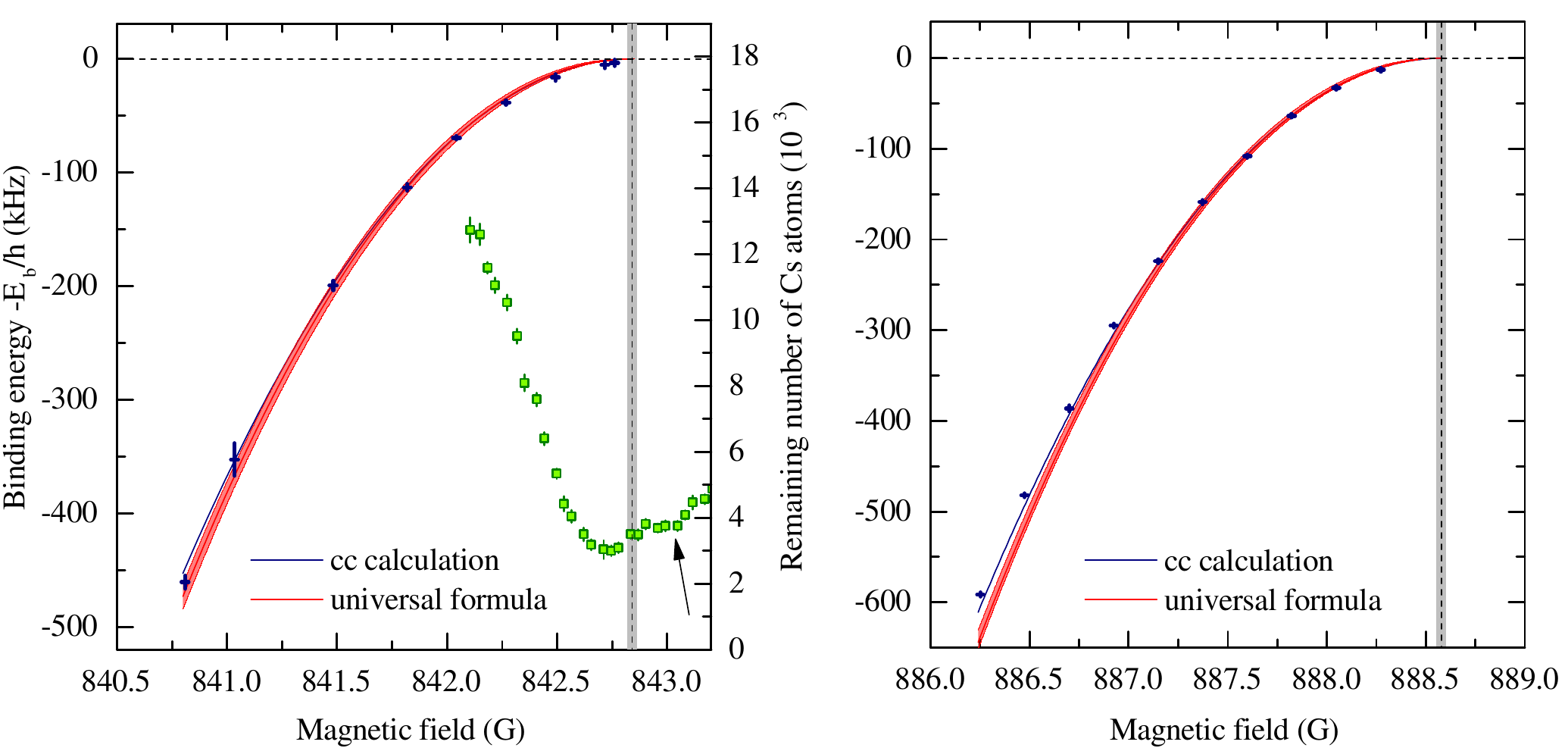}
 	\caption[843G binding energy]{Binding energies of LiCs Feshbach molecules and atom losses. The left and right panel correspond to the magnetic field regions near the 843 G and 889 G Feshbach resonance, respectively. The blue crosses display the dimer binding energy $ -E_b/\hbar $ that is extracted from a fit of Eq.~\ref{eq:rate_eq} to the rf association spectrum at the given magnetic field. The error bars represent one standard deviation of the total error, which results from statistical and systematic uncertainties. The blue and red solid lines show the calculated molecular state energies from the coupled-channels model and the universal binding energy $ E_b=\hbar^2/(2\mu a^2) $ with the resonance parameters from Table \ref{tab:sWaves}, respectively. The red shaded region corresponds to the uncertainty of the FR parameters. The green squares show the remaining Cs atom number for a corresponding atom loss measurement. Here the error corresponds to one standard error of the mean. The systematic magnetic field uncertainty for the atom loss measurements in this figure is 30 mG. The vertical dashed line displays the resonance pole position, and the gray shaded region corresponds to the uncertainty. The arrow in the left panel shows the position of the  second excited LiCsCs Efimov resonance with scattering length $a^{(2)}_- $. }
 	\label{fig:Eb_843G}
 \end{figure}

The binding energy of the Feshbach dimers at a given magnetic field is extracted by fitting Eq. \ref{eq:rate_eq} to the loss spectrum of Li atoms, as displayed in Fig. \ref{fig:843G_LossSpectrum}. We use $ E_b $, $ N_{Li}^{0} $, $ \gamma $, and $ C $ as free fitting parameters and set $ a^{\prime}=-28.5 \ a_0 $ \cite{Repp2013,Tung2013,Pires2014a}. Small variations in $ a^{\prime} $ that are of the order of a few of percent have affect the fitted binding energies on a permille level. The temperature of each species is determined in an independent measurement with identical trapping parameters and is kept fixed during the fit. To exclude systematic effects associated with the precise determination of relative temperature we verify that by increasing it by a factor of two the value of the fitted binding energy does not change by more than 1 kHz. By performing the measurements and the fitting procedure for different external magnetic fields, we record the binding energy dependence, which is displayed in Fig. \ref{fig:Eb_843G} for the two broadest FRs in the Li-Cs mixture.

The extracted binding energy can be affected by several other systematic effects. One of them is the mean-field shift, which starts to dominate in the regime where the scattering length is comparable to the interparticle spacing, i.e. $ n a^3 \sim 1 $. For our experimental densities of $ n \approx 10^{11}\ \mathrm{cm}^{-3} $ such shifts would become relevant at magnetic field regions with binding energies on the order of $ E_b \approx 1\ \mathrm{kHz}$, which is sufficiently far away from the region where the experiments were performed. Additionally, by changing the background Cs atom density we checked that the observed molecular association line shifts stay within the statistical uncertainties of the fit, and therefore we do not include the mean-field shift in the analysis. 

Another source of systematic resonance shifts is the confining optical dipole potential. The detuning of the dipole trap laser beam is large in comparison to the hyperfine splitting of the involved spin states, hence its created light shift is equal for both of them and can be neglected. However, the confining potential can contribute to the scattering state energy shift in two other ways. The first one is the confinement induced shift of the relative ground state energy for two colliding free atoms. Its magnitude can be calculated for two interacting particles in a cigar shaped harmonic trap \cite{Idziaszek2006}, and for our dipole trap geometry with aspect ratio $ \eta\approx 9 $ it yields 325 Hz. The second complication is the fact that, in general, the problem of two different atoms in a harmonic trap with unlike trapping frequencies does not separate into center-of-mass and relative coordinates. The magnitude of the shift of the associated lowest energy state in a Li-Cs mixture can be estimated for our trapping geometry and mass-ratio, and is on the order of 50 Hz \cite{Bertelsen2007}. Since the order of magnitude of these corrections is much smaller than the measured binding energies we neglect these effects in the model that we use to fit the data, however, we include them in the total systematic error budget. 

To obtain a complete set of Li-Cs s-wave FR properties we re-measure the positions of the narrow s-wave resonances in the two energetically lowest scattering channels up to 1000 G. For these measurements we reach roughly an order of magnitude lower relative kinetic energies than in the previous works \cite{Repp2013,Tung2013}. Their experimental positions $ B^e $ are determined with a Gaussian fit to each of the loss features in the magnetic field range where the line shape is approximately symmetric. The improved magnetic field stability and lower temperatures allows us to determine these resonance positions with a roughly five-fold better precision than in our previous measurement \cite{Repp2013}. These results are summarized in Table \ref{tab:sWaves}.

\subsection{Coupled-channels calculation}
To obtain an accurate mapping between the scattering length and the magnetic field that is independent of the employed analytical fitting model \cite{Julienne2014}, we analyze the data with the full cc calculation for the Li(2s)+Cs(6s) asymptote as in the previous work \cite{Repp2013,Pires2014a}. In short, the determination of the final resonance positions relies on the creation of accurate LiCs molecular potential curves for the electronic singlet $ X ^{1}\Sigma^+ $ and triplet $ a ^{3}\Sigma^+ $ ground states. The potentials are constructed in such a way that they simultaneously reproduce the binding energies of the Feshbach molecules, the refined s-wave FR positions $ B^e $ from atom loss experiments, and 6498 rovibrational transitions from laser-induced Fourier-transform spectroscopy \cite{Staanum2007}. We deduce the theoretical resonance positions $ B^t $ from the maxima of calculated two-body collision rates at the experimental kinetic energy. For the binding energies below the 843 G resonance we exclude the two data points with the smallest binding energies from the fit. Their rf association spectra, due to experimental limitations, already overlap with the Li free-free transition spectra,  which hinders a reliable extraction of the free-bound spectra (similar to the one in Fig. \ref{fig:843G_LossSpectrum}) for these respective magnetic field values. 

The results of the modeling are listed in Table \ref{tab:sWaves} as deviation  $\delta=B^{e}-B^{t} $ from the measured positions for the experimentally employed relative collision temperature $ T $ and drawn as solid lines in Fig. \ref{fig:Eb_843G}. These results provide almost an order of magnitude improvement over the previous determination of the FR positions through the trap-loss measurements \cite{Repp2013,Tung2013} and the rf spectroscopy \cite{Pires2014}, and they are consistent with the recent theoretical analysis \cite{Pires2014a}, if the differences in determining the resonance positions and experimental accuracy are taken into account.

\begin{table}
	\caption{\label{tab:sWaves}
		Positions of the LiCs s-wave Feshbach resonances. Unless specifically noted, the experimentally obtained resonance positions $ B^e $ are extracted by fitting a Gaussian profile to the loss spectra for the relative collision temperature $ T $, at which the measurements were made. The numbers in the brackets represent the total error that includes uncertainty of the magnetic field, and statistical and systematic errors of determining the position of the resonance. The results of the coupled-channels calculation $ B^t $ are given as deviations $\delta=B^{e}-B^{t} $ from the observations and show excellent agreement with the data. $ B_{FR} $, $\Delta$, and $ a_{bg} $ gives the fitted resonance pole position, width, and background scattering length, respectively, for the calculation with kinetic energy of 1 nK. 
}   % potential model used 87o 
\begin{indented}
	\lineup
	\item[]\begin{tabular}{@{}c|cll|lll}
		          \br 
			Entrance channel            & $ B^e $ (G) & $\delta$ (G) & $ T $ (nK) & $ B_{FR} $ (G) & $\Delta$ (G) & $ a_{bg} $ ($ a_0 $) \\
		\mr
		    Li$\left|1/2,-1/2\right\rangle $  &  816.128(20)  &-0.005        & 300        & 816.113    & -0.37        & -29.6                \\
		 $ \oplus \text{Cs} \left|3,+3\right\rangle$  &  888.595(16)*   & 0.002        & 100        & 888.578    & -57.45        & -29.6                \\
		                                              &  943.020(50)  & -0.033       & 400        & 943.033    & -4.22        & -29.6                \\
		\mr
		     Li$\left|1/2,+1/2\right\rangle $ &  842.845(16)*   & -0.000        & 100        & 842.829    & -58.21        & -29.4                \\
		 $ \oplus \text{Cs} \left|3,+3\right\rangle$  &  892.655(30)$ ^\dag $  & 0.005        & 100        & 892.629    & -4.55        & -29.4 \\ \br
	\end{tabular}
		\item[]* Extrapolated from rf association. The temperature shown is only used for the calculation of the scattering resonance and selected sufficiently low to reduce its influence to less than 5 mG. The error reflects the uncertainty of the field calibration.
		\item[]$ ^\dag $ This measurement was performed in a double-wavelength optical dipole trap with species selective optical potentials. Details will be given elsewhere.

\end{indented}
\end{table}

Finally, we characterize the resonance profiles by calculating the scattering length dependence on the magnetic field at a kinetic energy of 1 nK and fitting this dependence with the conventional functional form 
\begin{equation} \label{eq:a_B}
	a=a_{bg} \left(1-\frac{\Delta}{B-B_{FR}} -\ldots\right)
\end{equation}
with as many terms as there are resonances in the given channel. The resonance position $ B_{FR} $, its width $ \Delta $, and background scattering length $ a_{bg} $ are used as free fitting parameters, and they are given in Table \ref{tab:sWaves}. By including all observed resonances in a single fit of Eq.~\ref{eq:a_B} one removes a possible slope of the effective background scattering length resulting from neighboring resonances. We note that one could also fit the calculated profiles by a product of resonance functions for each resonance instead of the sum. This will result mainly in different values of $\Delta$, but as long as one is using only the derived functional values, the interpretation is consistent. The fitted values reproduce the calculated s-wave scattering length to better than 2\% in the entire magnetic field range between 500 G and 1000 G, which we also use for the re-evaluation of the Efimov resonance positions (see the next section). There is a slight difference between the two theoretically obtained resonance pole positions $ B^t $ and $ B_{FR} $. We strongly suspect that it originates from the different types of numerical calculations that were used to extract these parameters, however further investigation is necessary to find the exact reason behind this difference.  Therefore we estimate the error for the resonance pole positions from the systematic error from the magnetic field calibration and difference between the theoretical values, which yields $\pm$ 23 mG.

We use the resonance parameters that were obtained from the cc calculation to plot the simple single-channel formula $ E_b=\hbar^2/(2\mu a^2) $, which relates the universal dimer binding energy to the scattering length (see Fig.~\ref{fig:Eb_843G}). The Li-Cs characteristic van der Waals energy scale \cite{chin2010} is 157 MHz, thus the influence of the short range effects on the measured binding energies is minimal. This is reflected in the nearly ideal $ 1/a^2 $ scaling of the measured binding energies in this magnetic field range. Since the Li-Cs background scattering length $ a_{bg}\approx-29.5\, a_0$ is small and negative, we expect only very minor influence of the virtual state in this regime. This contributes to the simple situation where the two LiCs binding energies are well described with the universal relation and can be treated independently from other neighboring resonances in the same scattering channels. This is in contrast to more complicated situations, like the one in Cs atoms where FRs overlap \cite{Lange2009,Jachymski2013}.

Our determined position of the 843 G FR pole clearly deviates from the previously observed atom loss maximum \cite{Pires2014,Tung2014}, as can be seen in Fig.~\ref{fig:Eb_843G}. It also deviates from the result $ B_{FR}=842.75(3) $ G obtained by Tung~et~al.~\cite{Tung2014}, where exclusively atom loss measurements are used to infer the resonance pole position. This illustrates that the use of atom loss alone is questionable for a reliable determination of the FR pole position, especially if the resonance width is much larger than experimental uncertainties, as it is in the present case. The definition of the resonance pole position is based on pure two-body scattering whereas a number of different loss mechanisms may contribute to the total loss effect, the most prominent being the three-body collisions. The situation in the vicinity of the resonance's pole is complicated furthermore by the fact that not all of inelastic three-body collisions result in an immediate loss of the atoms from the trap. Contribution from other recombination processes, for example, weakly-bound dimer formation and subsequent atom-dimer recombination, should be considered, which may lead to increased loss away from the pole of the FR \cite{Zhang2011c}. In this case the maximum of total atom losses can be shifted with respect to the maximum in the corresponding three-body loss rate. The specific loss channels and the exact pathway of this decay in Li-Cs system still remain an open question, requiring a selective product state determination, which is not available at the present stage of our experiment. However, we expect that the shifts between the determined scattering pole positions and experimentally observed loss maxima can be explained or influenced by similar mechanisms as those discussed for other systems of ultracold gas mixtures \cite{Weber2008,Machtey2012,Zhang2011c, Khramov2012}.

\section{Scaling of LiCsCs Efimov resonances}

\begin{table}
	\caption{\label{tab:Efimov}
		Positions and scaling factors of the LiCsCs Efimov resonances near the 843 G Feshbach resonance. The error of the magnetic field, at which the resonances are found, represent the combined statistical and systematical error arising from the determination of the Efimov resonance positions and magnetic field uncertainty. For $ a^{(i)}_-$ and the scaling factors the uncertainty of the $ a(B) $ mapping through the Feshbach resonance parameters is added to the error. }
	
	\begin{indented}
		\lineup
		\item[]\begin{tabular}{l|ccc}
			\br
			Efimov state $ i $ & Magnetic field (G) & $ a^{(i)}_- $ $ (a_0) $ & $ a^{(i)}_-/a^{(i-1)}_- $ \\
			\mr
			0 (ground)         &     848.90(7)$ ^\dag $      &        -311(4)         &             -             \\
			1                         &     843.85(3)$ ^\dag $      &       -1710(70)        &          5.5(2)          \\
			2                         &     843.03(6)$ ^\dag $      &      -8540(2650)      &         5.0(1.5)         \\ \mr
			0                         &     848.55(12)*     &           -329(7)           &             -             \\
			1                         &     843.82(5)*      &           -1760(100)            &            5.3(3)            \\
			2                         &     842.97(4)*      &           -12200(4140)           &            6.9(2.3)            \\ \br
		\end{tabular} 
		\item[]$ ^\dag $ Ref. \cite{Pires2014}
		\item[]* Ref. \cite{Tung2014}

	\end{indented}

\end{table}

With the improved determination of the 843 G FR pole position we re-evaluate the scattering lengths and scaling factors of the previously observed Efimov resonances \cite{Pires2014,Tung2014}. In Table \ref{tab:Efimov} we summarize the reported magnetic field values and corresponding scattering lengths $a^{(i)}_- $, at which the $ i\mathrm{-th} $ three-body Efimov state merge with the scattering threshold. The extracted scaling factors of 5.5(2) and 5.3(3) for the first Efimov period are close to the expected value of 4.9 for a zero-temperature gas in the scaling limit ($ |a|\gg \bar{a} $) \cite{Braaten2006,DIncao2006a}, however they slightly deviate from the universal prediction. This is not surprising, since the assumption of the scaling limit is not strictly justified for the ground state Efimov resonance \cite{Pires2014}. Several theoretical studies have demonstrated that the ground Efimov state can be subject to large modifications due to short range physics \cite{Wang2012,Schmidt2012,Thogersen2008,Platter2009a,Naidon2011} and even three-body forces \cite{Dincao2009}. 
Furthermore, the recent study of the first excited-state resonance in Cs \cite{Huang2014,Wang2014a} and the new analysis of $^6$Li data \cite{Huang2014a} do not only give a hint to deviations from the universal scaling, but also to shifts of the ground-state resonances due to finite range effects. The exact origin of the above mentioned deviation in the Li-Cs system still remains an open question. The present analysis also does not consider Efimov resonance shifts arising from finite temperature. This is in contrast to the recent studies in the homonuclear gases of Cs and Li atoms \cite{Huang2014,Huang2014a,Rem2013}, which have shown that such effects need to be taken into account in order to accurately determinate the Efimov resonance positions. At the same time, the scaling factor of the second Efimov period lies well within the universal predictions, and experimental demonstration of deviations from the universal law, if any, will require access to lower temperatures and improved control over the magnetic field systematics.

\section{Conclusion}

In conclusion, we have precisely determined the resonance positions of all the s-wave FRs in the two energetically lowest Li-Cs scattering channels. The present work represents almost an order of magnitude improvement in accuracy and precision over the previous determination. It was achieved by performing rf association of colliding Li-Cs atom pairs into weakly bound dimers close to the broad FRs around 843 G and 889 G and complementary atom loss measurements of further three narrow s-wave resonances. Based on the measured magnetic field dependent binding energies and atom losses precise singlet and triplet molecular potential curves for the LiCs electronic ground state were constructed with the help of a cc calculation. The obtained potentials were used to map the Li-Cs scattering length on the external magnetic field. The precise resonance parameters will be pivotal for future experiments in the Li-Cs systems, where they can serve as a starting point for the preparation and investigation of ultracold polar molecules \cite{Carr2009,Dulieu2009,Jin2012}, the creation of strongly interacting mixtures and polarons \cite{Ferrier-Barbut2014,Kohstall2012,Koschorreck2012,Schirotzek2009}, and enable access to the Efimov physics in the universal regime \cite{Ferlaino2011,Huang2014,Pires2014}.

Additionally, the accurate knowledge of the Li-Cs FR parameters has allowed us to re-evaluate the positions of previously observed Efimov resonances. We have found a slight deviation from the predicted universal scaling factor for the first period of LiCsCs Efimov trimers, while the scaling factor of the second period is consistent with the universal law. This result is intriguing, since it approaches a regime where the applicability of few-body theories to a realistic mass-imbalanced system can be tested quantitatively. To fully enter it, however, lower temperatures will be necessary. Presently observed Efimov features may be significantly influenced by the short range interactions and the unitary limit, which both obscure clear observation of the second excited LiCsCs Efimov resonance and might have strong effect already on the first excited one. This may lead to shifts of the resonance positions and require more sophisticated models for accurate description of the three- and more-body physics near FRs.

\ack
We thank C. Greene, R. Grimm, S. Jochim, and G. Z\"urn for fruitful discussions. We gratefully acknowledge contributions to the experiment by C. Renner. This work is supported in part by the Heidelberg Center for Quantum Dynamics. R.P., S.H. and J.U. acknowledge support by the IMPRS-QD and J.U. by the DAAD. E.K. acknowledges support by the Baden-W\"urttemberg Stiftung and E.T. by the Volkswagen-Stiftung.

\section*{References}
\bibliographystyle{iopart-numMOD}
\bibliography{T:/quantdyn/Literature/Mixtures/Mixtures}% Produces the bibliography via BibTeX.

\providecommand{\newblock}{}
\begin{thebibliography}{10}
\expandafter\ifx\csname url\endcsname\relax
  \def\url#1{{\tt #1}}\fi
\expandafter\ifx\csname urlprefix\endcsname\relax\def\urlprefix{URL }\fi
\providecommand{\eprint}[2][]{\url{#2}}

\bibitem{Braaten2006}
Braaten E and Hammer H~W 2006 {\em Physics Reports\/} {\bf 428} 259 -- 390 ISSN
  0370-1573

\bibitem{Ferlaino2011}
Ferlaino F, Zenesini A, Berninger M, Huang B, N\"agerl H~C and Grimm R 2011
  {\em Few-Body Systems\/} {\bf 51}(2-4) 113--133 ISSN 0177-7963

\bibitem{Wang2013}
Wang Y, D'Incao J~P and Esry B~D 2013 Ultracold few-body systems {\em Advances
  in Atomic, Molecular, and Optical Physics\/} ({\em Advances In Atomic,
  Molecular, and Optical Physics\/} vol~62) ed Ennio~Arimondo P~R~B and Lin C~C
  (Academic Press) pp 1 -- 115

\bibitem{chin2010}
Chin C, Grimm R, Julienne P and Tiesinga E 2010 {\em Rev. Mod. Phys.\/} {\bf
  82}(2) 1225--1286

\bibitem{Koehler2006}
K\"ohler T, G\'oral K and Julienne P~S 2006 {\em Rev. Mod. Phys.\/} {\bf 78}(4)
  1311--1361

\bibitem{Weber2008}
Weber C, Barontini G, Catani J, Thalhammer G, Inguscio M and Minardi F 2008
  {\em Phys. Rev. A\/} {\bf 78}(6) 061601

\bibitem{Machtey2012}
Machtey O, Kessler D~A and Khaykovich L 2012 {\em Phys. Rev. Lett.\/} {\bf
  108}(13) 130403

\bibitem{Zhang2011c}
Zhang S and Ho T~L 2011 {\em New Journal of Physics\/} {\bf 13} 055003

\bibitem{Khramov2012}
Khramov A~Y, Hansen A~H, Jamison A~O, Dowd W~H and Gupta S 2012 {\em Phys. Rev.
  A\/} {\bf 86}(3) 032705

\bibitem{Klempt2008}
Klempt C, Henninger T, Topic O, Scherer M, Kattner L, Tiemann E, Ertmer W and
  Arlt J~J 2008 {\em Phys. Rev. A\/} {\bf 78}(6) 061602

\bibitem{Regal2003}
Regal C~A, Ticknor C, Bohn J~L and Jin D~S 2003 {\em Nature\/} {\bf 424} 47--50
  ISSN 0028-0836

\bibitem{Bartenstein2005}
Bartenstein M, Altmeyer A, Riedl S, Geursen R, Jochim S, Chin C, Denschlag J~H,
  Grimm R, Simoni A, Tiesinga E, Williams C~J and Julienne P~S 2005 {\em Phys.
  Rev. Lett.\/} {\bf 94}(10) 103201

\bibitem{Ospelkaus2006}
Ospelkaus C, Ospelkaus S, Humbert L, Ernst P, Sengstock K and Bongs K 2006 {\em
  Phys. Rev. Lett.\/} {\bf 97}(12) 120402

\bibitem{Wu2012}
Wu C~H, Park J~W, Ahmadi P, Will S and Zwierlein M~W 2012 {\em Phys. Rev.
  Lett.\/} {\bf 109}(8) 085301

\bibitem{Zuern2013}
Z\"{u}rn G, Lompe T, Wenz A~N, Jochim S, Julienne P~S and Hutson J~M 2013 {\em
  Phys. Rev. Lett.\/} {\bf 110}(13) 135301

\bibitem{Thompson2005}
Thompson S~T, Hodby E and Wieman C~E 2005 {\em Phys. Rev. Lett.\/} {\bf 95}(19)
  190404

\bibitem{Papp2006}
Papp S~B and Wieman C~E 2006 {\em Phys. Rev. Lett.\/} {\bf 97}(18) 180404

\bibitem{Lange2009}
Lange A~D, Pilch K, Prantner A, Ferlaino F, Engeser B, N\"agerl H~C, Grimm R
  and Chin C 2009 {\em Phys. Rev. A\/} {\bf 79}(1) 013622

\bibitem{Gross2010}
Gross N, Shotan Z, Kokkelmans S and Khaykovich L 2010 {\em Phys. Rev. Lett.\/}
  {\bf 105}(10) 103203

\bibitem{Berninger2013}
Berninger M, Zenesini A, Huang B, Harm W, N\"agerl H~C, Ferlaino F, Grimm R,
  Julienne P~S and Hutson J~M 2013 {\em Phys. Rev. A\/} {\bf 87}(3) 032517

\bibitem{Dyke2013}
Dyke P, Pollack S~E and Hulet R~G 2013 {\em Phys. Rev. A\/} {\bf 88}(2) 023625

\bibitem{Gribakin1993}
Gribakin G~F and Flambaum V~V 1993 {\em Phys. Rev. A\/} {\bf 48}(1) 546--553

\bibitem{Repp2013}
Repp M, Pires R, Ulmanis J, Heck R, Kuhnle E~D, Weidem\"uller M and Tiemann E
  2013 {\em Phys. Rev. A\/} {\bf 87}(1) 010701

\bibitem{Tung2013}
Tung S~K, Parker C, Johansen J, Chin C, Wang Y and Julienne P~S 2013 {\em Phys.
  Rev. A\/} {\bf 87}(1) 010702

\bibitem{Pires2014a}
Pires R, Repp M, Ulmanis J, Kuhnle E~D, Weidem\"uller M, Tiecke T~G, Greene
  C~H, Ruzic B~P, Bohn J~L and Tiemann E 2014 {\em Phys. Rev. A\/} {\bf 90}(1)
  012710

\bibitem{Pires2014}
Pires R, Ulmanis J, H\"afner S, Repp M, Arias A, Kuhnle E~D and Weidem\"uller M
  2014 {\em Phys. Rev. Lett.\/} {\bf 112}(25) 250404

\bibitem{Tung2014}
Tung S~K, Jim\'enez-Garc\'ia K, Johansen J, Parker C~V and Chin C 2014 {\em
  Phys. Rev. Lett.\/} {\bf 113}(24) 240402

\bibitem{DIncao2006a}
D'Incao J~P and Esry B~D 2006 {\em Phys. Rev. A\/} {\bf 73}(3) 030703

\bibitem{Kerman2000}
Kerman A~J, Vuleti\ifmmode~\acute{c}\else \'{c}\fi{} V, Chin C and Chu S 2000
  {\em Phys. Rev. Lett.\/} {\bf 84}(3) 439--442

\bibitem{Chin2005a}
Chin C and Julienne P~S 2005 {\em Phys. Rev. A\/} {\bf 71}(1) 012713

\bibitem{Idziaszek2006}
Idziaszek Z and Calarco T 2006 {\em Phys. Rev. A\/} {\bf 74}(2) 022712

\bibitem{Bertelsen2007}
Bertelsen J~F and M\o{}lmer K 2007 {\em Phys. Rev. A\/} {\bf 76}(4) 043615

\bibitem{Julienne2014}
Julienne P~S and Hutson J~M 2014 {\em Phys. Rev. A\/} {\bf 89}(5) 052715

\bibitem{Staanum2007}
Staanum P, Pashov A, Kn\"ockel H and Tiemann E 2007 {\em Phys. Rev. A\/} {\bf
  75}(4) 042513

\bibitem{Jachymski2013}
Jachymski K and Julienne P~S 2013 {\em Phys. Rev. A\/} {\bf 88}(5) 052701

\bibitem{Wang2012}
Wang J, D'Incao J~P, Esry B~D and Greene C~H 2012 {\em Phys. Rev. Lett.\/} {\bf
  108}(26) 263001

\bibitem{Schmidt2012}
Schmidt R, Rath S and Zwerger W 2012 {\em The European Physical Journal B\/}
  {\bf 85} 1--6 ISSN 1434-6028

\bibitem{Thogersen2008}
Th\o{}gersen M, Fedorov D~V and Jensen A~S 2008 {\em EPL (Europhysics
  Letters)\/} {\bf 83} 30012

\bibitem{Platter2009a}
Platter L, Ji C and Phillips D~R 2009 {\em Phys. Rev. A\/} {\bf 79}(2) 022702

\bibitem{Naidon2011}
Naidon P and Ueda M 2011 {\em Comptes Rendus Physique\/} {\bf 12} 13 -- 26 ISSN
  1631-0705 few body problem Problème à petit nombre de corps

\bibitem{Dincao2009}
D'Incao J~P, Greene C~H and Esry B~D 2009 {\em Journal of Physics B: Atomic,
  Molecular and Optical Physics\/} {\bf 42} 044016

\bibitem{Huang2014}
Huang B, Sidorenkov L~A, Grimm R and Hutson J~M 2014 {\em Phys. Rev. Lett.\/}
  {\bf 112}(19) 190401

\bibitem{Wang2014a}
Wang Y and Julienne P~S 2014 {\em Nat Phys\/} {\bf 10} 768–773 ISSN 1745-2481

\bibitem{Huang2014a}
Huang B, O'Hara K~M, Grimm R, Hutson J~M and Petrov D~S 2014 {\em Phys. Rev.
  A\/} {\bf 90}(4) 043636

\bibitem{Rem2013}
Rem B~S, Grier A~T, Ferrier-Barbut I, Eismann U, Langen T, Navon N, Khaykovich
  L, Werner F, Petrov D~S, Chevy F and Salomon C 2013 {\em Phys. Rev. Lett.\/}
  {\bf 110}(16) 163202

\bibitem{Carr2009}
Carr L~D, DeMille D, Krems R~V and Ye J 2009 {\em New Journal of Physics\/}
  {\bf 11} 055049

\bibitem{Dulieu2009}
Dulieu O and Gabbanini C 2009 {\em Reports on Progress in Physics\/} {\bf 72}
  086401

\bibitem{Jin2012}
Jin D~S and Ye J 2012 {\em Chemical Reviews\/} {\bf 112} 4801--4802 pMID:
  22967213

\bibitem{Ferrier-Barbut2014}
Ferrier-Barbut I, Delehaye M, Laurent S, Grier A~T, Pierce M, Rem B~S, Chevy F
  and Salomon C 2014 {\em Science\/} {\bf 345} 1035

\bibitem{Kohstall2012}
Kohstall C, Zaccanti M, Jag M, Trenkwalder A, Massignan P, Bruun G~M, Schreck F
  and Grimm R 2012 {\em Nature\/} {\bf 485} 615--618 ISSN 0028-0836

\bibitem{Koschorreck2012}
Koschorreck M, Pertot D, Vogt E, Fr\"ohlich B, Feld M and K\"ohl M 2012 {\em
  Nature\/} {\bf 485} 619--622 ISSN 0028-0836

\bibitem{Schirotzek2009}
Schirotzek A, Wu C~H, Sommer A and Zwierlein M~W 2009 {\em Phys. Rev. Lett.\/}
  {\bf 102}(23) 230402

\end{thebibliography}
\end{document}